\documentclass[10pt,twocolumn,letterpaper]{article}

\usepackage{cvpr}              

\usepackage{graphicx}
\usepackage{amsmath}
\usepackage{amssymb}
\usepackage{booktabs}

\usepackage[pagebackref,breaklinks,colorlinks]{hyperref}

\usepackage[capitalize]{cleveref}
\crefname{section}{Sec.}{Secs.}
\Crefname{section}{Section}{Sections}
\Crefname{table}{Table}{Tables}
\crefname{table}{Tab.}{Tabs.}

\def\confName{CVPR}
\def\confYear{2022}

\begin{document}

\title{
Leveraging Video Coding Knowledge for Deep Video Enhancement}
\author{
Thong Bach\\
{\tt\small thongbtqm@gmail.com}
\and
Thuong Nguyen Canh\\
{\tt\small ngcthuong@ids.osaka-u.ac.jp} 
\and
Van-Quang Nguyen\\
{\tt\small quang@vision.is.tohoku.ac.jp}
}
\maketitle
\begin{abstract}
Recent advancements in deep learning techniques have significantly improved the quality of compressed videos. However, previous approaches have not fully exploited the motion characteristics of compressed videos, such as the drastic change in motion between video contents and the hierarchical coding structure of the compressed video. This study proposes a novel framework that leverages the low-delay configuration of video compression to enhance the existing state-of-the-art method, BasicVSR++. We incorporate a context-adaptive video fusion method to enhance the final quality of compressed videos. The proposed approach has been evaluated in the NTIRE22 challenge, a benchmark for video restoration and enhancement, and achieved improvements in both quantitative metrics and visual quality compared to the previous method.
\end{abstract}

\section{Introduction}
\label{sec:intro}
With the increasing demand for high-quality video transmission over the Internet, video compression has become essential to efficiently transmit videos over limited bandwidth. It has driven the development of video compression standards such as H.265/HEVC \cite{hevc}, and beyond. However, compressed videos suffer unavoidable compression artifacts. As a result, there is a growing interest in the research community in enhancing the quality of compressed videos.

Several studies have proposed methods to improve the quality of individual frames in videos \cite{yang2018enhancing,wang2017novel} as well as leveraging temporal information between frames \cite{yang2019quality,guan2019mfqe,deng2020spatio,yang2020learning,wang2020multi,huo2021recurrent}. Most existing methods focus on the architecture design of the model, which typically involves (i) designing the backbone extraction module using CNNs or Transformers, (ii) designing the propagation module to effectively capture information flow between frames, and (iii) designing the enhancement module as a post-processing step to improve the quality of the output video. However, there is often a little emphasis on incorporating prior knowledge from video content such as motion information as well as the compression algorithm. This represents an untapped potential for improving the overall quality of the video compression process.

In this study, we propose several methods to enhance the performance of BasicVSR++, a state-of-the-art video super-resolution method \cite{chan2021basicvsr++}. Our approach begins by examining BasicVSR++'s performance with varying numbers of input frames, taking into account the motion information of the content. As compressed video uses the HEVC low-delay configuration, the first frame (also known as the Intra frame) has significantly higher quality than the others. To take advantage of this, we train a separate network called Intra frame BasicVSR++ to improve the quality of the first frame. Finally, we introduce an adaptive mechanism that combines multiple reconstructed instances with different input sequence lengths to obtain the final enhanced output. 

The experiments demonstrate that the proposed framework not only leverages the low-delay configuration of video compression but also incorporates context-adaptive video fusion to enhance the final quality of compressed videos. These results demonstrate the potential of incorporating domain-specific knowledge into deep learning models for advancing the state-of-the-art in compressed video quality enhancement.

\begin{figure*}[t]
    \centering
    \includegraphics[scale=0.65]{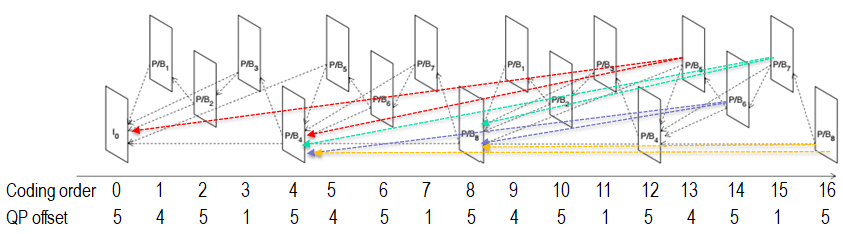}
    \caption{
    Low-delay configuration of HEVC with a group of picture size of 4, where the configuration has only one intra frame at the 0-th index and a repeated group structure for 4 frames. Full references are only available from the 13-th frame.
    }
    \label{fig:low_delay}
\end{figure*}

\section{Performance Analysis of BasicVSR++}
\begin{figure*}[t]
    \centering
    \includegraphics[scale=0.65]{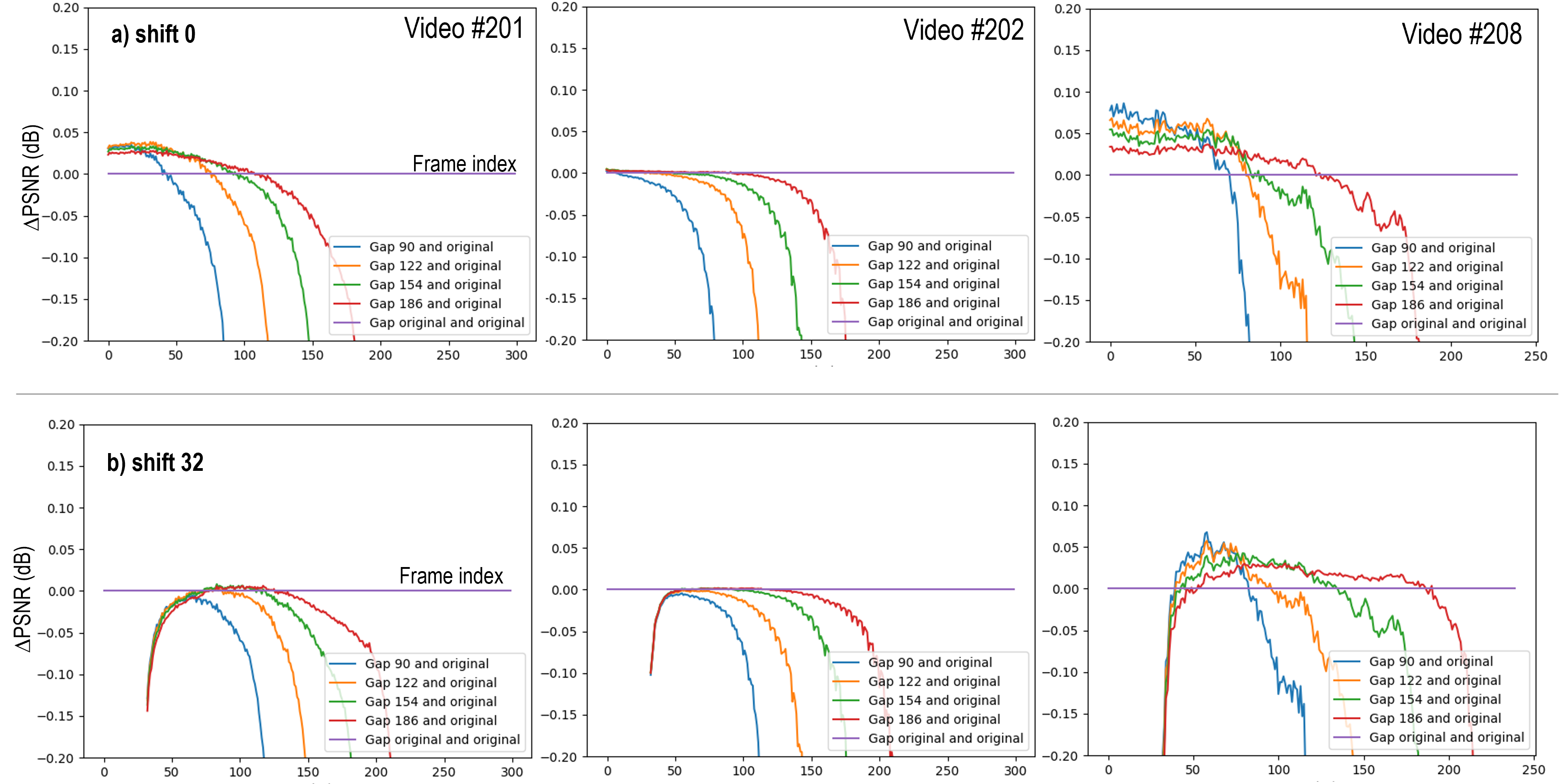}
    \caption{Performance variation of BasicVSR++ with respect to the number of input frames. The per-frame PSNR difference between the outputs of trimmed videos with different gaps ($90, 122, 154, 186$) and the output of the original video is shown as $\Delta$ PSNR. The first and second rows display the results with start frame $0$ and $32$, respectively.}
    \label{fig:gap_visual}
\end{figure*}

BasicVSR++ \cite{chan2021basicvsr++} is a state-of-the-art video super-resolution method that enhances video quality through a combination of frame propagation and alignment techniques. While BasicVSR++ has shown impressive results in enhancing video quality, it does not take into account the unique characteristics of compressed video.

In a compressed video \cite{hevc}, a frame can be an intra or inter frame. Intra frame compression only uses information from the current image, while inter-frame compression utilizes information from previously encoded frames to reduce redundancy. The NTIRE22 challenge encoded video using a low-delay configuration \cite{hevc}, as shown in Figure \ref{fig:low_delay}, with a group of pictures of 4. This results in the compressed video having only one intra frame at a significantly higher quality than other inter-frames. As a result, the quality of frames in the compressed video is highly varied, but BasicVSR++ does not take this into account. In practice, the entire video is fed as input to the model during testing, which may not be the optimal choice.


We investigate the effect of varying input video frame lengths on the performance of BasicVSR++. As in Fig. \ref{fig:gap_visual}-a, the pre-trained network's performance varies significantly depending on the frame length, with shorter frame length demonstrating higher performance for the first few frames.
Interestingly, we also noticed that this performance phenomenon did not occur when the input of the trimmed video started from the 32nd frame, i.e., when the intra frame is not included. This finding suggests that the network is better able to exploit the high performance of the intra frame with a smaller frame length. For later frames, our experiments demonstrated that using the full frame length performed better thanks to the temporal dependency in both backward and forward directions. In contrast, trimmed video inputs have limited backward and forward dependency, resulting in lower performance for later frames.

This observation suggests that the optimal choice of input frame length may vary depending on the temporal characteristics of the video content. The use of shorter frame lengths may be more effective for the early frames, while longer frame lengths may be more suitable for later frames that rely on both backward and forward dependencies. Additionally, the effectiveness of the backward and forward dependencies may be limited to a certain temporal range, as seen in Video 208 at shift 32. Examining 40 test sequences 201-240 in the NTIRE22 dataset, we observe that the first 64 frames of enhanced outputs with the video inputs trimmed at the length of 186 yields the best performance.

\begin{figure*}[t]
    \centering
    \includegraphics[scale=0.65]{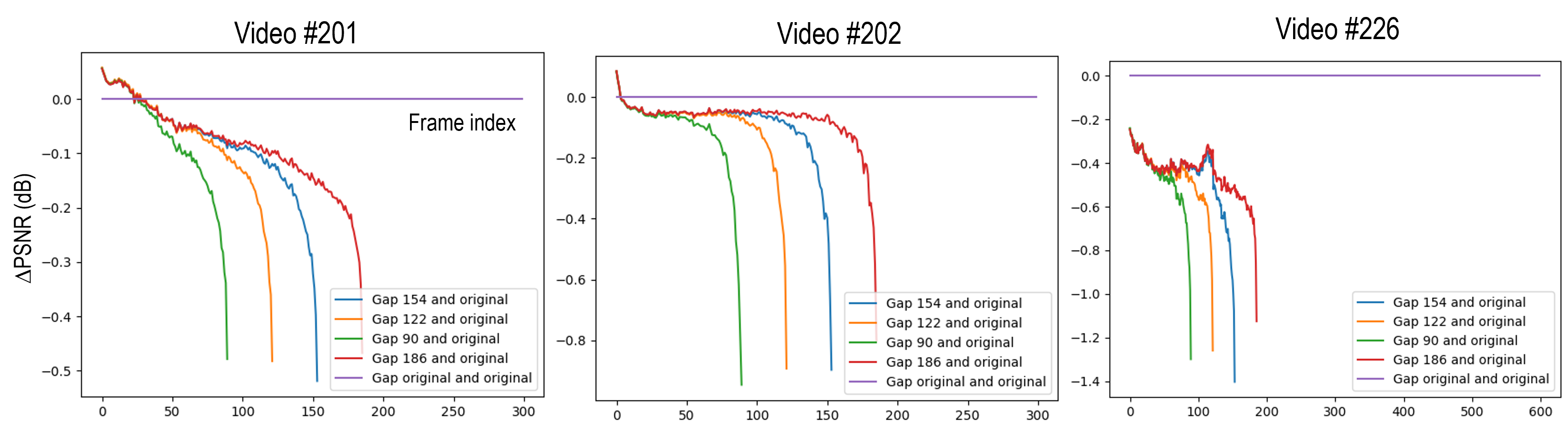}
    \caption{Performance of the fine-tuned Intra frame BasicVSR++ without the Adaptive Context-Aware Fusion mechanism.}
    \label{fig:i_frame}
\end{figure*}


\section{Proposed Method}
\subsection{Intra frame BasicVSR++}

\begin{figure*}[ht]f
    \centering
    \includegraphics[scale=0.45]{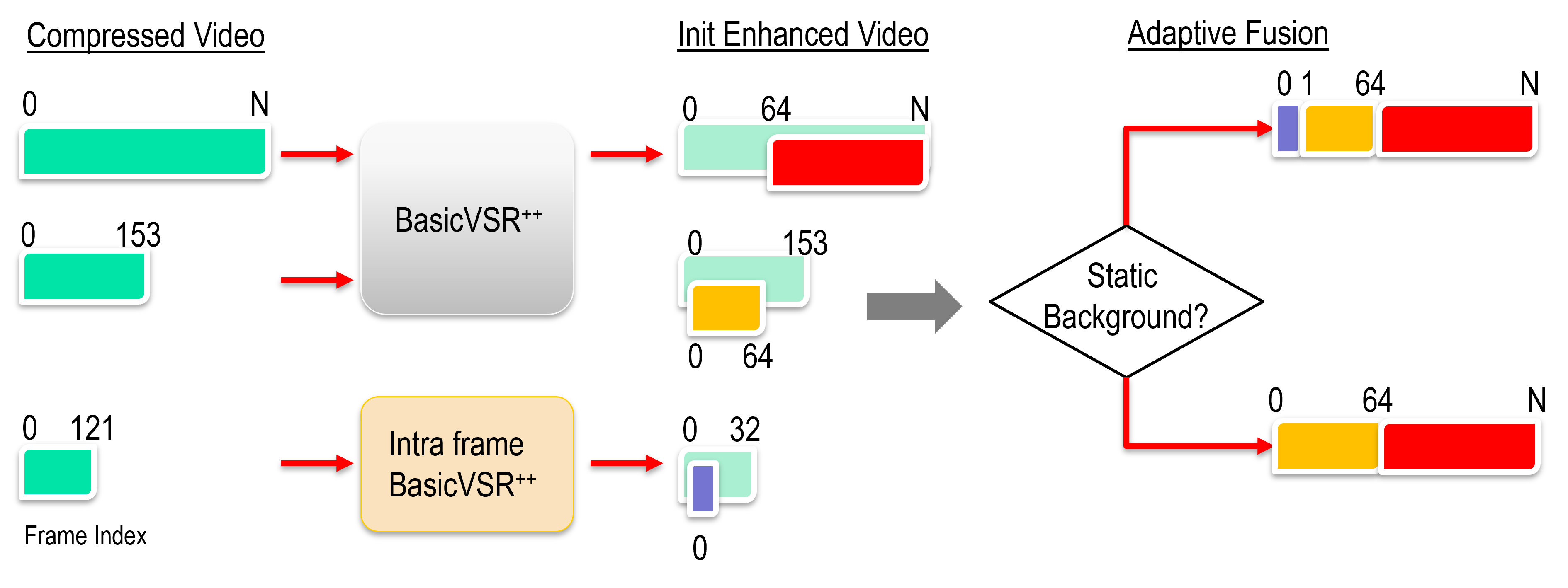}
    \caption{General framework of the adaptive context-aware mechanism for video fusion. The mechanism detects static video content using the gradient of the average frame.}
    \label{fig:general}
\end{figure*}

To further leverage the superior quality of the intra frame, we introduce a new network called Intra rame BasicVSR++. In order to do so, we created a new intra frame video dataset by cutting the original videos into multiple non-overlapping 30-frame segments and encoding them using HEVC with the same low-delay configuration as the NTIRE22 dataset. The resulting dataset contains videos with only one intra frame and multiple inter-frames. We then divided the compressed videos into training and testing datasets for network training. By utilizing this configuration, we ensured that the intra frame is of higher quality than the inter-frames, reflecting the reality of compressed videos.

During training, we fine-tuned the Intra frame BasicVSR++ network by using the intra frame as the first frame of each segment. This allowed the network to learn to enhance the high-quality intra frame more effectively, resulting in a more accurate and efficient network. In this way, the Intra frame BasicVSR++ network is designed to specifically target the improvement of the first frame's quality, while BasicVSR++ improves the quality of the entire video.


As shown in Fig.\ref{fig:i_frame}, we observed that the Intra frame BasicVSR++ network improves the performance of the intra frame in most cases, such as videos 201 and 202. However, this improvement is not universal and may not always hold for sequences with high frame rates and slow motion, such as video 226. This may be due to the limited amount of information in the intra frame in such cases, and the network may struggle to extract and propagate useful features. Nonetheless, these results suggest that the intra frame BasicVSR++ network can be an effective tool in improving the performance of compressed video enhancement for most types of content.

\subsection{Adaptive Context-Aware Fusion}
From our analysis, it is evident that BasicVSR++ benefits from adaptive input frame length and can be further improved by Intra frame BasicVSR++. However, it is also important to note that this improvement is not applicable to all types of video content. In order to address this issue, we propose a heuristic that separates cases where the frame rate is high and there is slow motion. This is achieved by comparing the gradient of the average frame with a given threshold. By using this threshold, we can determine if the video contains a high frame rate and slow motion, and take appropriate measures to optimize the final performance.
Firstly, the average frame is obtained as follows:

\begin{equation}
\bar{f} = \sum_{i}^{i \times m < N} f_{i\times m} ,
\end{equation}
where $m$ is a scaling factor that is proportional to the input video frame rate. Specifically, we set $m=4$ for videos with a frame rate less than 30fps and $m=8$ for those with a frame rate greater than 30fps. Next, we compare the gradient of the average frame to a given threshold:

\begin{equation}
\nabla (\bar{f}) = ||\nabla_x (\bar{f})|| + ||\nabla_y (\bar{f})|| < \tau,
\end{equation}
where $\nabla_x$ and $\nabla_y$ denote the gradient in the horizontal and vertical directions of a given frame $f$, and $\tau$ is a threshold of value 2300. The value of $\tau$ can be normalized based on the number of pixels, as shown in Fig. \ref{fig:general}.


We propose a novel video fusion mechanism called Adaptive Context-Aware Fusion, as shown in Fig. \ref{fig:general}. The method involves enhancing an input video using three different approaches: full BasicVSR++, short BasicVSR++ with the first 154 frames, and the first 122 frames by Intra frame BasicVSR++. Depending on the current content of the video, an adaptive fusion technique is performed to select either the first frame from short BasicVSR++ or Intra frame BasicVSR++. For the subsequent 63 frames, we select frames from the short BasicVSR++ set, and the remaining frames are extracted from the full BasicVSR++ set. 

\subsection{Loss Function}

In order to fine-tune the BasicVSR++ and Intra frame BasicVSR++ networks, we used a weighted sum of three loss components: (i) Charbonnier loss \cite{lai2018fast}, (ii) total variation (TV) loss, and (iii) temporal gradient (TG) loss. The temporal gradient loss captures the difference between the ground truth and the output temporal sequence by calculating the loss between two consecutive frames in each sequence. To calculate the temporal gradient loss, we subtract two consecutive frames in each sequence to obtain the temporal gradient sequence. 

The hyperparameters of the loss weights were optimized using grid search to find the best values. The final loss function is given by:

\begin{equation}
\begin{aligned}
\mathcal{L}_{\mathrm{Final}} = \mathcal{L}_{\mathrm{Char}} +1e\times 10^{-3}*\mathcal{L}_{\mathrm{TG}}  \\\ + 1e\times 10^{-4}*\mathcal{L}_{\mathrm{TV}},
\end{aligned}
\end{equation}
where $L_{Char}$ is the Charbonnier loss, $L_{TV}$ is the total variation loss, $L_{TG}$ is the temporal gradient loss.

\section{Experiments}
\subsection{Datasets}
For the NTIRE 2022 Challenge, we used the original LDV dataset \cite{yang2021ntire} that consists of 240 videos as our primary training set. To increase our training data, we also utilized the LDV 2.0 dataset, which contains an additional 90 videos. We split the LDV 2.0 videos into six sets, each containing 15 videos. Two of these sets were used as our validation and test sets, respectively. In splitting the videos, we aimed to maintain the diversity of the videos in each set, in terms of content, frame rate, and other factors, as similar as possible. All videos in the LDV and LDV 2.0 datasets, as well as the splits for the NTIRE 2021 and NTIRE 2022 Challenges, are publicly available at \url{https://github.com/RenYang-home/LDV_dataset}.

\subsection{Training Details}
We employed the Adam optimizer \cite{kingma2014adam} with a learning rate of $2 \times 10^{-5}$ and utilized the Cosine Restart scheduler \cite{loshchilov2016sgdr} with a period of 10,000 iterations. To ensure a stable optimization process, we linearly increased the learning rate for the first 10\% of iterations.

\subsubsection{Fine-tuning BasicVSR++}
Due to computational limitations, we fine-tuned only the upsample layer of the pre-trained BasicVSR++ network. We found that increasing the input frame length from 30 to 60 frames led to a 0.03 dB improvement in model performance. The network was fine-tuned for 50,000 iterations.

\subsubsection{Fine-tuning Intra frame BasicVSR++}

We create a new dataset is created by trimming the original videos into multiple segments, each of which consists of 30 frames without overlapping frames. The video segments are encoded using HEVC at a low-delay profile resulting in approximately 16,000 training samples. For the Intra frame BasicVSR++ network, only the segments from the first 200 videos are used to train the model, and the last 40 videos are used for testing.

\begin{table*}[t]
\footnotesize
  \centering
  \caption{The PSNR results on the test set for our method and the compared method, BasicVSR++. Values are given in dB.}
    \resizebox{\textwidth}{!}{
\begin{tabular}{lcccccccccccccccc}
\toprule
Method \textbackslash Video ID  & 1  & 2 & 3 & 4 & 5 & 6 & 7 & 8 & 9 & 10 & 11 & 12 & 13 & 14 & 15  &Avg.  \\
 \midrule
BasicVSR++ & 33.73 & 32.42&  31.22 & 36.75&  32.16&  30.57 & 30.41&  33.99&  31.68 & 36.20&  27.06 & 23.93&  30.24 & 32.94&  31.20&  31.63\\
Ours (OCL-VCE) & 33.81 &  32.54 & 31.30&  36.82 & 32.28&  30.59&  30.47 & 34.07&  31.75&  36.28 & 27.08&  24.00&  30.37&  33.01&  31.25&  31.71\\
\bottomrule
\end{tabular}
}
\label{tab:res}
\end{table*}

\subsection{Ensembling with Test Time Augmentation}

In our study, we perform Ensembling with test time augmentation (TTA) by generating eight input variations by flipping and rotating the input sequences in the spatial dimension. We then use our proposed framework shown in Fig. \ref{fig:general} to enhance each variation, and post-process the corresponding output by flipping and rotating it back to its original location. Finally, we obtain the final output by averaging the outputs of all variations. This approach helps to reduce the impact of input variability and improve the overall performance of the model.

\subsection{Experimental Results}
We participated in the NTIRE22 challenge as team \textbf{OCL-VCE} and submitted our results on the test set. In Table \ref{tab:res}, we present the performance of our method and BasicVSR++ on the test set. Our method achieved a higher PSNR score of 31.71 dB compared to 31.63 dB obtained by BasicVSR++.

We further evaluated our framework on the validation set, which consists of 10 videos. Using the pre-trained BasicVSR++ and Intra frame BasicVSR++ models and applying test-time augmentation (TTA), our framework achieved a PSNR score of 32.12 dB, surpassing the 31.84 dB score achieved without TTA.

In addition, without finetuning, with and without TTA, our framework achieved a PSNR score of 32.02 dB and 31.86 dB, respectively, improving the baseline by 0.06 dB and 0.02 dB, respectively. Our results demonstrate the effectiveness of our proposed Adaptive Context-Aware Fusion framework in enhancing the quality of low-delay compressed videos.

\section{Conclusion}
In conclusion, this paper proposes a novel method that leverages the unique characteristics of low-delay video compression algorithms to improve the quality of compressed videos using deep learning techniques. By incorporating this prior knowledge into the state-of-the-art method, BasicVSR++, we achieve a significant improvement in performance over existing methods. Our experimental results on the NTIRE22 challenge validate the effectiveness of our proposed method. This work underscores the importance of incorporating video compression knowledge into deep learning models to further enhance their performance and enable real-world applications.
{\small
\bibliographystyle{IEEEtran}

\bibliography{egbib}
}




\end{document}